\documentclass[fleqn,usenatbib]{mnras}

\usepackage[T1]{fontenc}

\DeclareRobustCommand{\VAN}[3]{#2}
\let\VANthebibliography\thebibliography
\def\thebibliography{\DeclareRobustCommand{\VAN}[3]{##3}\VANthebibliography}

\usepackage{graphicx}	
\usepackage{amsmath}	
\usepackage{amssymb}	
\usepackage{caption}
\usepackage{subcaption}
\usepackage{multirow}
\usepackage{multicol}
\usepackage{pdflscape}
\usepackage{float}
\usepackage{newtxtext,newtxmath}

\title[Epicyclic frequencies of a heterogeneous spheroid]{Epicyclic frequencies of spheroidal stars with non-uniform density.}

\author[D. A. Bollimpalli]{
D. A. Bollimpalli,$^{1,2}$\thanks{E-mail: deepika@mpa-garching.mpg.de}\\
$^{1}$Max Planck Institute for Astrophysics, Karl-Schwarzschild-Str. 1, D-85741 Garching, Germany.\\
$^{2}$Nicolaus Copernicus Astronomical Center, ul. Bartycka 18, PL 00-716 Warsaw, Poland.\\
}

\date{Accepted XXX. Received YYY; in original form ZZZ}

\pubyear{2015}

\begin{document}
\label{firstpage}
\pagerange{\pageref{firstpage}--\pageref{lastpage}}
\maketitle

\begin{abstract}
We consider the gravitational potential of a rotating star with non-uniform density to derive the orbital and epicyclic frequencies of the particles orbiting the star. We assume that the star is composed of concentric spheroids of constant density, with a global power-law distribution of density inside the star. At the lowest order approximation, we recover the known result for the Maclaurin spheroid that the maximum in the radial epicyclic frequency occurs at $r=\sqrt{2}ae$, for eccentricities $\geq 1/\sqrt{2}$. We find that the nature of these characteristic frequencies differs based on the geometry of the rotating star. For an oblate spheroid, the orbits resemble retrograde-Kerr orbits and the location of the radial epicyclic maximum approaches the stellar surface as the density variation inside the star becomes steeper. On the contrary, orbits around a prolate spheroid resemble prograde-Kerr orbits, but the marginally stable orbit does not exist for prolate-shaped stars. The orbital frequency is larger (smaller) than the Keplerian value for an oblate (prolate) star with the equality attained as $e\rightarrow0$ or $r\rightarrow\infty$. The radial profiles of the angular velocity and the angular momentum allow for a stable accreting disc around any nature of oblate/prolate spheroid.
\end{abstract}



\section{Introduction}
The study of orbital motions has been of great importance in various astrophysical scenarios. In many cases, the gravitational potential is axisymmetric in nature, which confines the motion to near-circular orbits. Deviations from such near-circular orbits are termed epicycles. To the lowest order, they correspond to the harmonic motion in the equatorial plane/ meridional plane of the gravitating body, characterized by the radial/ vertical epicyclic frequency, respectively. Together with the orbital frequency, both epicyclic frequencies characterize the motion of an orbiting particle. For example, stars orbiting in galactic discs or particles/fluid orbiting a star can undergo such epicyclic motions and they can be well formulated in Newtonian gravity.

For a spherically symmetric potential in Newtonian gravity (like $1/r$), both the radial epicyclic ($\omega_r$) and vertical epicyclic ($\omega_z$) frequencies are the same as the orbital frequency ($\Omega_0$). In most cases Newton’s theory of gravity is sufficient to study the orbital motions, however, near strong gravitational sources like compact objects, relativistic effects are significant and should be accounted for. For a non-rotating, spherically symmetric object, orbits in Newtonian and Einstein's gravity differ by three-fold with the latter case having: (i) a marginally stable orbit (where the specific angular momentum has a minimum), (ii) a maximum in the radial epicyclic frequency, and (iii) a broken degeneracy among the orbital and epicyclic frequencies. 

The existence of a marginally stable orbit (at $r=r_{\rm ms}$) is in fact an important feature of general relativity. Though the marginally stable orbit does not exist for the $1/r$ Newtonian potential, it has been shown in \citet{WR2013}, that for a rapidly rotating Maclaurin spheroid with eccentricity larger than $0.834583178$, a marginally stable orbit can exist outside the stellar surface even in Newtonian gravity, and the radial epicyclic frequency exhibits a maximum over an even larger range of eccentricities. Further, the orbits around Maclaurin spheroid in Newtonian gravity have $\omega_r^2<\Omega_0^2<\omega_z^2$ with the degeneracy in epicyclic and orbital frequencies broken by the rotation \citep{Oglivie2008, WR2013}. The above orbital properties of the Maclaurin spheroid resemble the retrograde orbits in the Kerr metric. Note that the direction of orbits is irrelevant in Newtonian gravity while the properties of the orbits differ for prograde and retrograde orbits in the Kerr metric. 

The significance of epicyclic oscillations is well appreciated in Einstein's gravity as the trapped oscillation modes\footnote{Certain properties of the radial distribution of the epicyclic frequencies in general relativity allows for the trapping of different oscillation modes in the inner regions of the relativistic accretion discs.} in discs offer a favourable explanation for the quasi-periodic oscillations (QPOs) observed in the neutron star and black hole binaries \citep{W1999, K01, RM06}. 
Similar quasi-coherent variability is observed on timescales of seconds to tens of minutes in cataclysmic variables, characterized as dwarf nova oscillations (DNOs) and QPOs \citep{WR72, PRN77}.
As relativity is of little importance in white dwarfs, the favoured explanation for the DNOs relates to the interaction between the accretion disc and the white dwarf's magnetic field \citep{WWII02, KLA05}, while the accretion disc is believed to be the source of the QPOs \citep{CMSC85, CHV2000, WWI02}. For a detailed review of the observations and favoured models for DNOs and QPOs in white dwarfs, refer to \citet{W04}. More observations and timing analysis in the near future will certainly help in better understanding the origins of this rapid variability, however at the same time developing theoretical models that could provide possible explanations for such and similar variability is also quite important.

The analytical calculations of orbital motions around a uniform density, Maclaurin spheroid showed that the properties of the orbits resemble retrograde Kerr orbits, thus opening a new channel for QPO models \citep{ABG02, WR2013}. Studies suggest that stars with non-uniform density under hydromagnetic equilibrium tend to distort into spheroids based on the geometrical conditions; global poloidal magnetic fields tend to distort the star into an oblate spheroid, while toroidal magnetic fields tend to deform the star into prolate spheroid \citep{CF1953, W61}. If the star is distorted into a spheroid, then we must expect differences in brightness between polar and equatorial regions. Furthermore, non-uniform density distribution alters the potential if the object is a spheroid/ellipsoid with a certain eccentricity and this could have important dynamical effects on the motion of the particles around the star. In this paper, we, therefore, aim to study the orbital dynamics for heterogeneous spheroids (both prolate and oblate in shape) and compare the properties of the orbits to that of the Maclaurin spheroids.

A brief outline of the paper is as follows: In section~\ref{section2} we introduce the gravitational potential of a heterogeneous ellipsoid (both prolate and oblate) at a given radius, and use this to derive the general expressions for the orbital and the epicyclic frequencies in section~\ref{section3}. Here, we only consider the orbits in the equatorial plane of the star. In section~\ref{section4}, we discuss how the properties of the orbits differ for a prolate and oblate spheroid, and how they vary based on the density distribution of the star. We also briefly discuss the observational relevance of these frequencies in this section. Finally, we summarize our results in section~\ref{section5}.

\section{Potential of a heterogeneous ellipsoid}
\label{section2}
In this section, we revisit the potential of a heterogeneous ellipsoid derived in \cite{Chandrasekhar1969}, and adapt it to the physical problem we are interested in by making certain assumptions. Consider a solid heterogeneous ellipsoid with semi-axes $a_i$ for $i=1,2,3$. We assume that the ellipsoid is composed of ellipsoidal strata of equal density that are similar to and concentric with the bounding ellipsoid (the outermost stratum). Under this assumption, the strata of equal density can be simply treated as ellipsoids with semi-axis $ma_i$, where $m$ is a parameter ranging from zero at the centre to one on the surface of the ellipsoid. For such an ellipsoid, its density $\rho$, can be expressed as a function of the square of $m$. 

The potential of such heterogeneous ellipsoid at an external point $x_i$ is given as \footnote{Here $x_1, x_2, x_3$ are the axes of an orthogonal coordinate system} [see Theorem 10, Chapter~3 of \cite{Chandrasekhar1969}]
\begin{equation}
\Phi = -\pi G a_1 a_2 a_3 \int_\lambda^\infty \dfrac{du}{\Delta} \int_{m^2(u)}^1 {\rm d}m^2 \rho(m^2),
\label{eq.genpot}
\end{equation}
where $\lambda$ is the ellipsoidal coordinate of $x_i$ with respect to the bounding ellipsoid and so is the largest positive root of the equation,
\begin{equation}
    \sum_{i=1}^3 \dfrac{x_i^2}{a_i^2+\lambda}=1.
    \label{eq.lambda}
\end{equation}
Here $u$ is a variable of integration that ranges over $\lambda$ to $\infty$ for the given integral. The function $m^2(u)$ is defined as
\begin{equation}
\sum_{i=1}^3 \dfrac{x_i^2}{a_i^2+u}=m^2.
\label{eq.m}
\end{equation}
It is easy to see that as $u\rightarrow \lambda$, $m\rightarrow 1$ corresponding to the bounding ellipsoid and as $u\rightarrow \infty$, $m\rightarrow 0$ which is at the centre of the ellipsoid. Also, 
\begin{equation}
\Delta = [(a_1^2+u)(a_2^2+u)(a_3^2+u)]^{\frac{1}{2}}.
\label{eq.delta}
\end{equation}

A specific form for the density distribution needs to be assumed for evaluating the integral in equation~\ref{eq.genpot}, and we choose the following form:
\begin{equation}
\rho(m^2) = \rho_0(1-m^2)^n, 
\label{eq.rho}
\end{equation}
where $n$ can be any positive number and $\rho_0$ is a constant, corresponding to the central density value. Substituting for this in equation~\ref{eq.genpot}, we arrive at
\begin{eqnarray}
\Phi = -\pi G a_1 a_2 a_3 \dfrac{\rho_0}{n+1} \int_\lambda^\infty \dfrac{du}{\Delta} \left(1-\sum\limits_{i=1}^3 \dfrac{x_i^2}{a_i^2+u}\right)^{n+1},
\label{eq.denpot}
\end{eqnarray}
Our interest lies in computing the epicyclic frequencies for rotating stars which can be well approximated with spheroids ($a_1=a_2 =a$ and $a_3 = b$). The potential for heterogeneous spheroid, when worked out in cylindrical coordinates, is given by
\begin{eqnarray}
\Phi(r,z) = -\pi G a^2b \dfrac{\rho_0}{n+1} \int_\lambda^\infty \dfrac{du}{\Delta} \left(1-\dfrac{r^2}{a^2+u}-\dfrac{z^2}{b^2+u}\right)^{n+1}.
\label{eq.potential}
\end{eqnarray}
Depending on whether the spheroid is oblate or prolate, we take $b = a(1-e^2)^{1/2}$ or $a = b(1-e^2)^{1/2}$, respectively, where $0\leq e \leq 1$ is the ellipticity/eccentricity. 
\section{Epicyclic frequencies}
\label{section3}
Consider the motion of a fluid element orbiting the central rotating star under the axisymmetric potential defined in equation~\ref{eq.potential}. In the absence of any additional external force (other than gravity), the orbital angular velocity, $\Omega_0$, of the fluid element is simply
\begin{equation}
\Omega_0^2(r) = \dfrac{1}{r} \dfrac{\partial \Phi(r,z)}{\partial r}\bigg\vert_{z=0}
\label{eq.omega_def}
\end{equation}
Note that $u$ is independent of the distance coordinates, $r$ and $z$, while $\lambda$ is a function of $r$. Therefore, following the Leibniz integral rule\footnote{$\frac{\mathrm{d}}{\mathrm{d}x} \left (\int_{g(x)}^{h(x)}f(x,t)\,\mathrm{d}t \right) = f(x,h(x))\cdot h'(x) - f(x,g(x))\cdot g'(x) + \int_{g(x)}^{h(x)} \frac{\partial}{\partial x}f(x,t)\; \mathrm{d}t$}, and the fact that $\Phi(r,z,\lambda) = 0$ by the definition of $\lambda$ (see eq.~\ref{eq.lambda}), we arrive at
\begin{equation}
\begin{split}
\dfrac{\Omega_0^2}{2\pi G\rho_0} =& a^3(1-e^2)^{\zeta} \\
& \times\int_\lambda^\infty \dfrac{{\rm d}u}{[a^2+u]^2[a^2(1-e^2)^{2\zeta}+u]^{1/2}}\left(1-\dfrac{r^2}{a^2+u}\right)^n,
\end{split}
\label{eq.omega_int}
\end{equation}
where $\zeta$ is $1/2$ and $-1/2$ for oblate and prolate spheroids, respectively.

With the help of the binomial expansion, the above expression can be rewritten as a series in $r^2$, 
\begin{eqnarray}
\dfrac{\Omega_0^2}{2\pi G\rho_0} =(1-e^2)^{\zeta} e^{-3}\left[\sum_{j=0}^n (-1)^j \binom nj\alpha_j \left(\dfrac{r}{ae}\right)^{2j}\right],
\label{eq.omega_bin}
\end{eqnarray}
where $\alpha_j$ ( for $j = 0,1,2,3,...$) are the coefficients of expansion given by
\begin{eqnarray}
\alpha_j = \int_\lambda^\infty \dfrac{(ae)^{2j+3}}{[a^2+u]^{j+2}[a^2(1-e^2)^{2\zeta}+u]^{\frac{1}{2}}}\,{\rm d}u
\label{eq.alpha_coeff}
\end{eqnarray}
The above integral can be simplified by using the parametrization $u = (a^2 e^2\csc^2 \theta - a^2)(1-e^2)^{\zeta-1/2}$, which gives
\begin{eqnarray}
  \alpha_j=\begin{cases}
    2\int_0^{\vartheta_{1}}(\sin \theta)^{2j+2}\,{\rm d}\theta, & \text{if $\zeta = 1/2$}.\\
    2(1-e^2)^{j+3/2}\int_0^{\vartheta_{2}}(\tan \theta)^{2j+2}\sec \theta\,{\rm d}\theta, & \text{if $\zeta = -1/2$}.
  \end{cases}
\end{eqnarray}
where $\vartheta_1 = \sin^{-1}[ae/r]$ and $\vartheta_2 = \tan^{-1}[ae/(r\sqrt{1-e^2})]$ for $\zeta = 1/2$ and $\zeta = -1/2$, respectively. For $j=0$, we have 
\begin{eqnarray}
  \alpha_0 =\begin{cases}
   \vartheta_{1} -\dfrac{1}{2}\sin 2\vartheta_{1}, \\ \hspace{5cm} \text{if $\zeta = 1/2$}.\\

    (1-e^2)^{3/2}[\tan \vartheta_{2} \sec \vartheta_{2} - \log(\tan \vartheta_{2}+\sec \vartheta_{2})],\\ \hspace{5cm}\text{if $\zeta = -1/2$}.
  \end{cases}
\end{eqnarray}

Knowing $\alpha_0$, one can calculate the rest coefficients by using the following recurrence relation: 
\begin{equation}
\alpha_{j+1} = \begin{cases}
\dfrac{1}{(j+2)}\left[\left(j+\dfrac{3}{2}\right) \alpha_j -\cos \vartheta_1\,(\sin \vartheta_1)^{2j+3}\right],\\ \hspace{5cm} \text{if $\zeta = 1/2$}.\\
\\
\dfrac{-(1-e^2)}{(j+2)}\left[\left(j+\dfrac{3}{2}\right) \alpha_j -(1-e^2)^{j+3/2}\sec \vartheta_2\,(\tan \vartheta_2)^{2j+3}\right],\\ 
 \hspace{5cm}\text{if $\zeta = -1/2$}.\\
\end{cases}
\label{eq.recurr}
\end{equation}

Equation~\ref{eq.omega_bin} gives the expression for the square of the angular velocity of a fluid element orbiting the star in an equatorial plane. If the particle is subjected to an infinitesimally small perturbation from its orbital motion while still maintaining the same angular momentum, it results in the oscillation of the fluid element around the original circular orbit due to the restoring nature of gravity and centrifugal forces. The frequency at which the particle oscillates is termed the epicyclic frequency and for pure radial oscillations, the radial epicyclic frequency is given by
\begin{equation}
\omega_r^2= \dfrac{\partial^2 \Phi_{\rm eff}(r,z)}{\partial r^2}\bigg\vert_{z=0}=2\Omega_0\left(2\Omega_0+r\dfrac{d\Omega_0}{dr}\right).
\label{eq.omegar}
\end{equation}
Here, $\Phi_{\rm eff}(r,z) \equiv \Phi(r,z)+l^2/(2r^2)$ is the effective potential of a particle whose specific angular momentum is $l\, (\equiv \Omega_0 r^2)$. Upon doing a little algebra, we arrive at the expression for $\omega_{\rm r}^2$ given by, 
\begin{equation}
\begin{split}
\dfrac{\omega_r^2}{2\pi G\rho_0} = (1-e^2)^{\zeta}e^{-3}\sum_{j=0}^{n} (-1)^j& \binom nj \left(\dfrac{r}{ae}\right)^{2j} \\
& \times \left[(2j+4) \alpha_j +r \dfrac{{\rm d} \alpha_j}{{\rm d} r} \right] ,
\end{split}
\label{eq:omegar_exp}
\end{equation}
where the second term in the parenthesis can be simplified as,
\begin{equation}
    r\dfrac{{\rm d}\alpha_j}{{\rm d}r} =  \begin{cases}
 -2\tan \vartheta_1(\sin \vartheta_1)^{2j+2}, & \text{if $\zeta = 1/2$}.\\
 -2(1-e^2)^{j+3/2}\sin \vartheta_2(\tan \vartheta_2)^{2j+2}, & \text{if $\zeta = -1/2$}.
\end{cases}
\end{equation}

Similar to the radial epicyclic frequency, we can also derive the vertical epicyclic frequency given as 
\begin{eqnarray}
\omega_z^2 &=&  \dfrac{\partial^2 \Phi_{\rm eff}(r,z)}{\partial z^2}\bigg\vert_{z=0},\nonumber\\
\Rightarrow \dfrac{\omega_z^2}{2\pi G \rho_0}&=& (1-e^2)^{\zeta}e^{-3}\left[\sum_{j=0}^{n}(-1)^j \binom{n}{j}\beta_j \left(\dfrac{r}{ae}\right)^{2j} \right],
\label{eq.omegaz}
\end{eqnarray}
where
\begin{eqnarray}
\begin{split}
\beta_j &= \int_\lambda^\infty \dfrac{(ae)^{2j+3}\,{\rm d}u}{[a^2+u]^{j+1}[a^2(1-e^2)^{2\zeta}+u]^{\frac{3}{2}}} \\
&= -2(j+1)\alpha_j-\dfrac{{\rm d} \alpha_j}{{\rm d} r}r.
\end{split}
\label{10}
\end{eqnarray}

Equations~\ref{eq.omega_bin}, \ref{eq:omegar_exp}, and \ref{eq.omegaz} give the analytical expressions for the orbital frequency, radial and vertical epicyclic frequencies, respectively in the potential of a heterogeneous spheroid. In the following section, we will discuss the properties of these frequencies, how they differ for the prolate and oblate spheroids and their possible relevance to observations.
\begin{figure}
\begin{subfigure}[t]{0.5\textwidth}
         \centering
         \includegraphics[width=\textwidth]{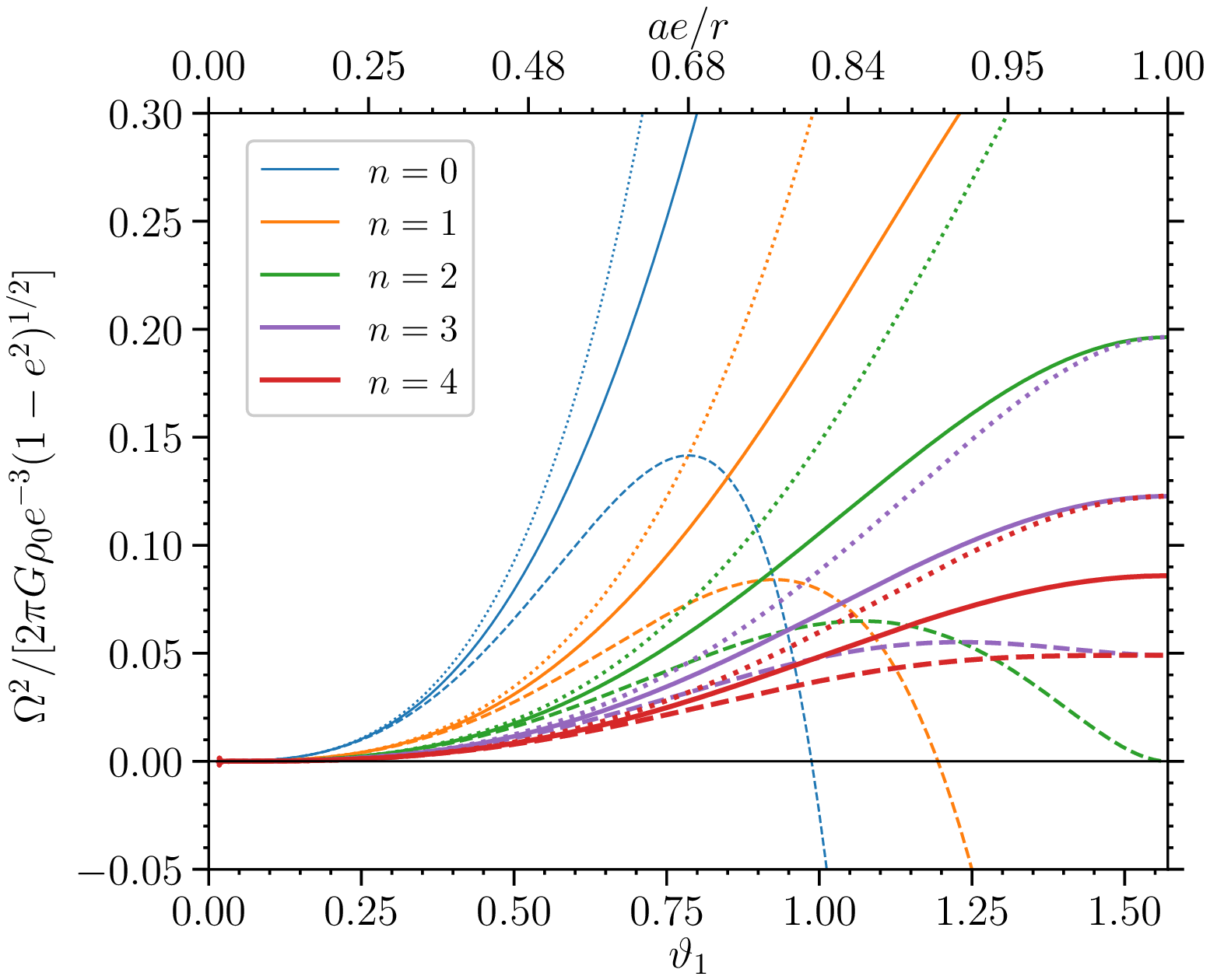}
\end{subfigure}
\begin{subfigure}[t]{0.5\textwidth}
         \centering
         \includegraphics[width=\textwidth]{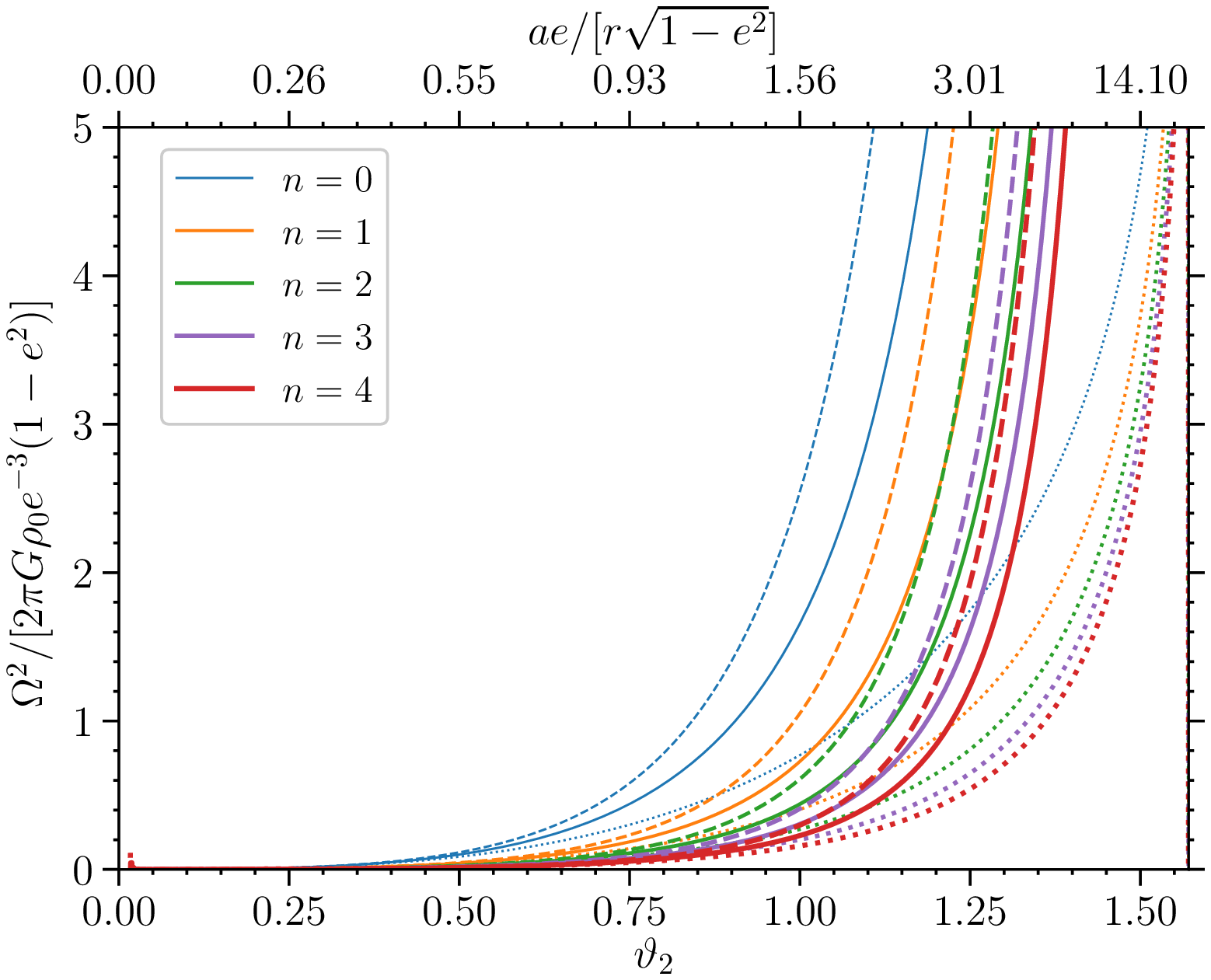}
\end{subfigure}
\caption{Plot of squares of angular velocity (solid curves), radial epicyclic frequency (dashed curves) and vertical epicyclic frequency (dotted curves) as a function of $\vartheta_1$ ($\vartheta_2$) for oblate (prolate) spheroid in the top (bottom) panel for $n = 0,1,2,3,4$ represented by the curves with increasing line thickness, respectively. The frequencies are scaled by a normalization factor as mentioned in the y-axis label.}
\label{fig:freq}
\end{figure}
\begin{figure*}
\begin{subfigure}[b]{0.45\textwidth}
         \centering
         \subcaption{e=0.5, oblate spheroid}
         \includegraphics[width=\textwidth]{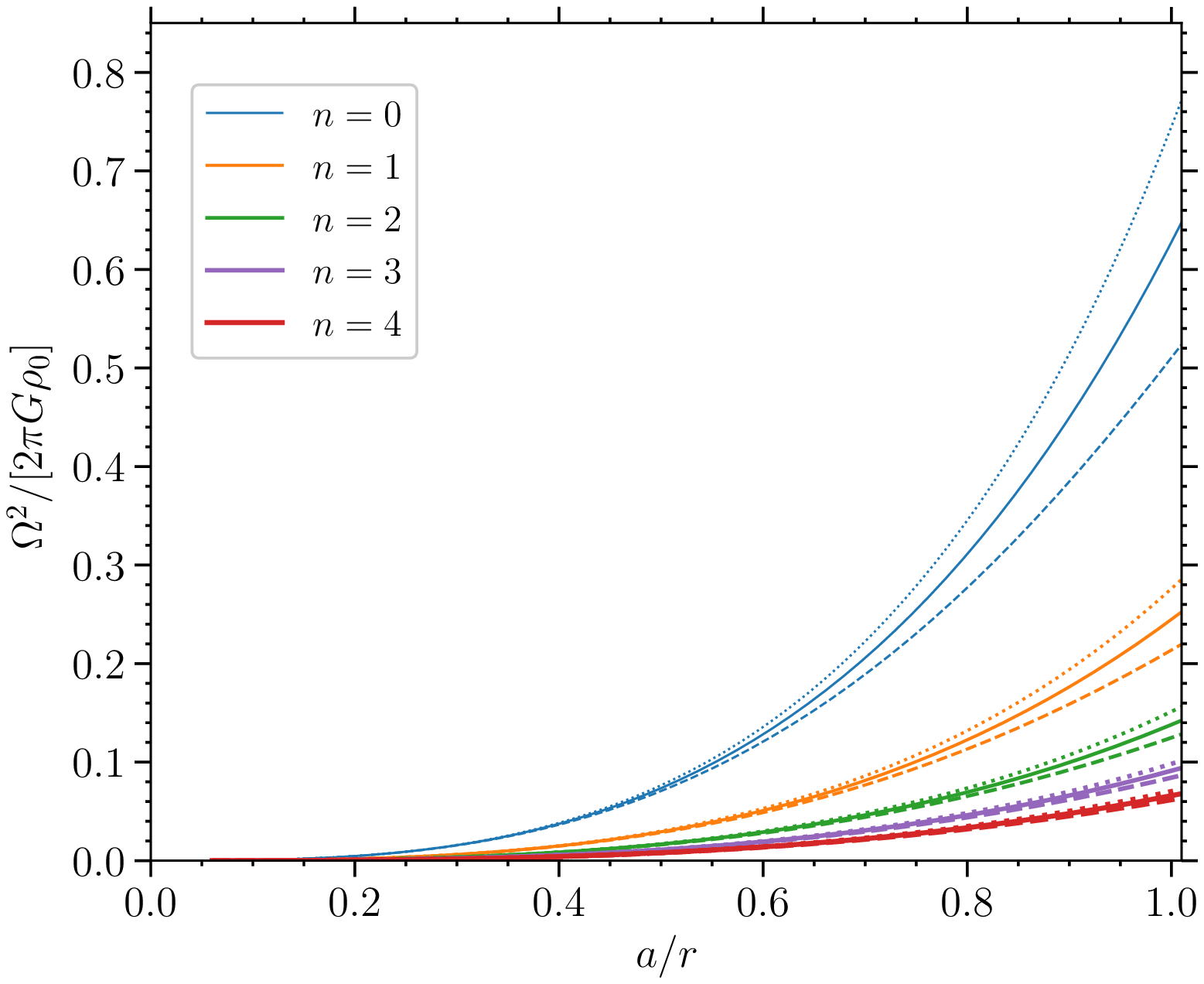}
\end{subfigure}
\begin{subfigure}[b]{0.45\textwidth}
         \centering
         \subcaption{e=0.9, oblate spheroid}
         \includegraphics[width=\textwidth]{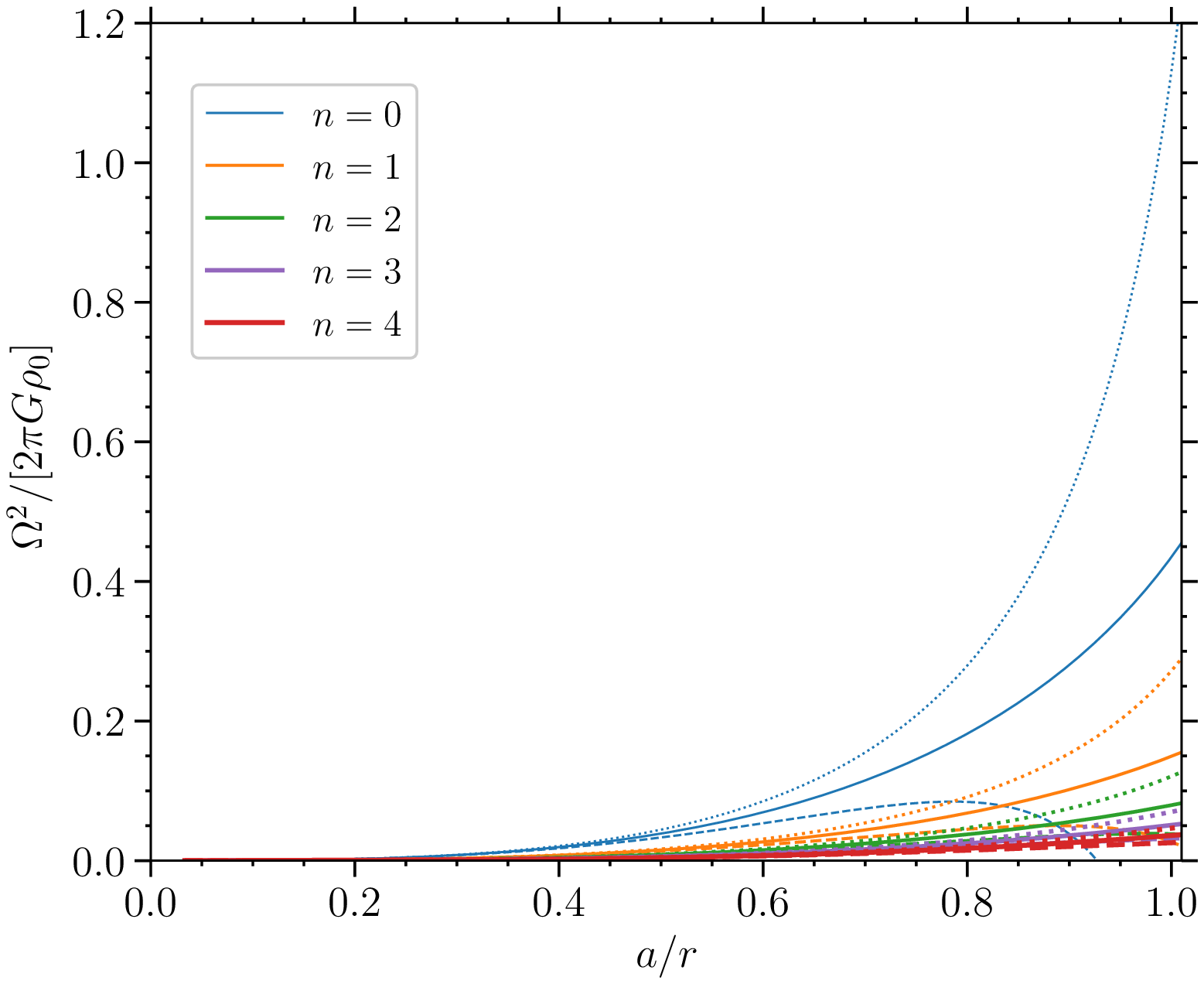}
\end{subfigure}
\\
\begin{subfigure}[b]{0.45\textwidth}
         \centering
         \subcaption{e=0.5, prolate spheroid}
         \includegraphics[width=\textwidth]{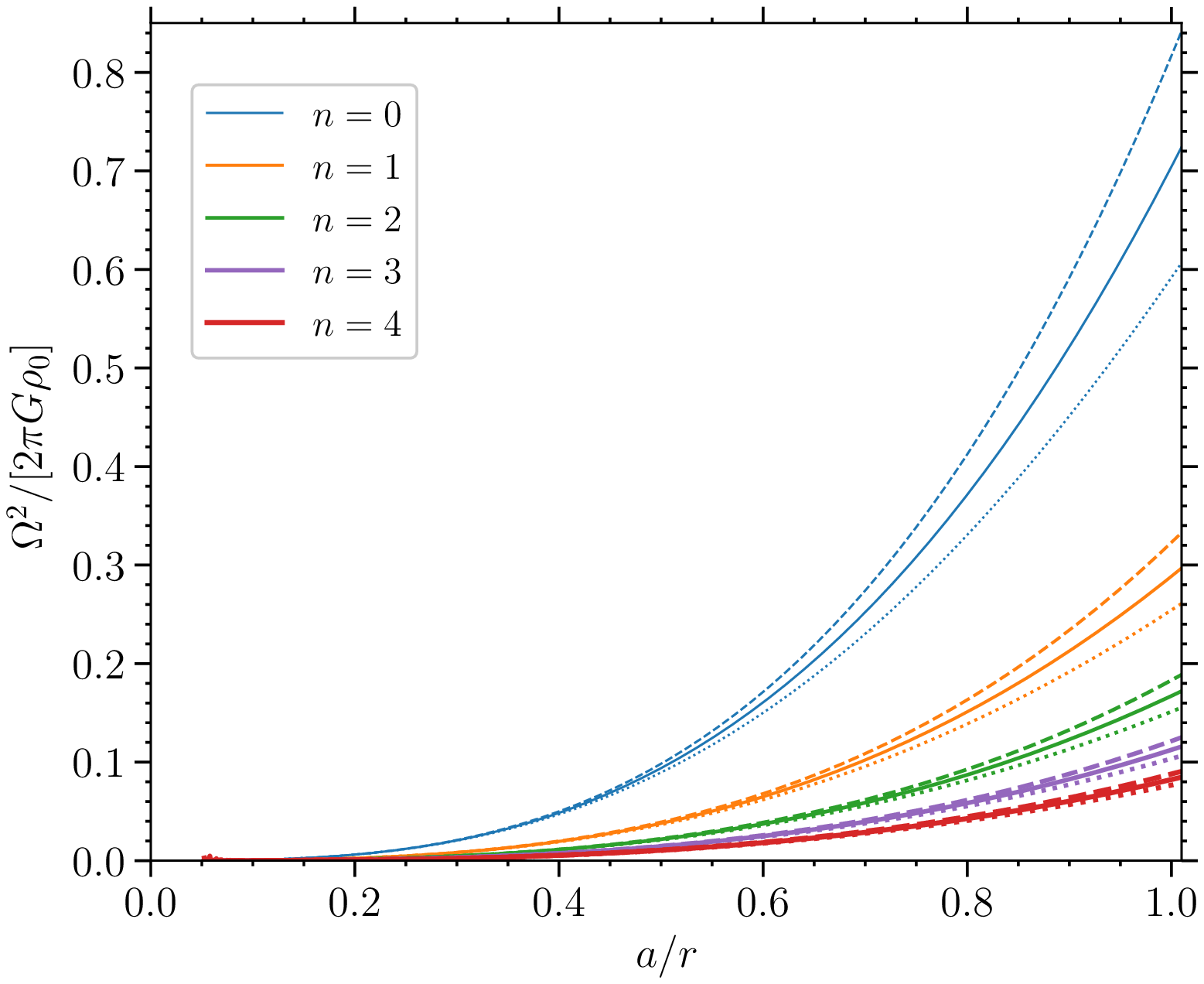}
\end{subfigure}
\begin{subfigure}[b]{0.45\textwidth}
         \centering
         \subcaption{e=0.9, prolate spheroid}
         \includegraphics[width=\textwidth]{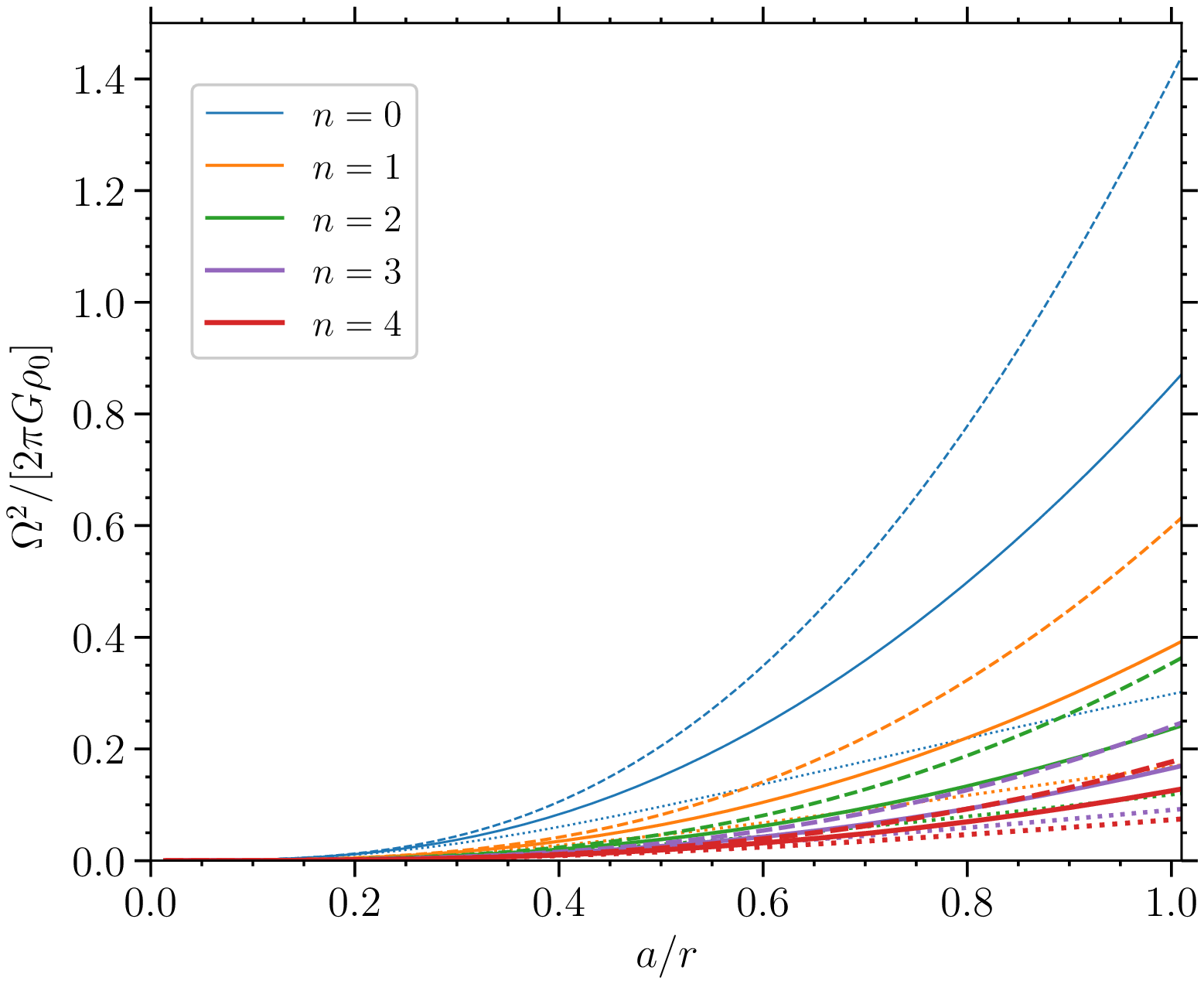}
\end{subfigure}
\caption{Radial profiles of squares of angular velocity (solid curves), radial epicyclic frequency (dashed curves) and vertical epicyclic frequency (dotted curves) for $e=0.5$ (left panels) and $e=0.9$ (right panels) are plotted for different $n$, with $n = 0,1,2,3,4$ represented by the curves with increasing line thickness, respectively. The top panel is for the oblate spheroid and the bottom panel is for the prolate spheroid configuration.}
\label{fig:freq_e0p5_e0p9}
\end{figure*}
\section{Properties of the epicyclic frequencies and their observational relevance}
\label{section4}
For orbits around an axisymmetric body, the radial and vertical epicyclic frequencies satisfy the relation,  
\begin{eqnarray}
\omega_r^2+\omega_z^2 = 2\Omega_0^2.
\label{eq:relation}
\end{eqnarray}
This relation is a direct result from the Laplace equation for any Newtonian potential in vacuum \citep{Shu1984}. For $n=0$, equation~\ref{eq.rho} implies constant density, and we recover the expressions for the epicyclic frequencies of a Maclaurin spheroid \citep{WR2013} by simply setting $n=0$ in the equations~\ref{eq.omega_bin},~\ref{eq:omegar_exp} and \ref{eq.omegaz}. 

The top panel of Fig.~\ref{fig:freq} shows the dependence of the angular and epicyclic frequencies on the compactified radius ($ae/r$) for orbits around an oblate spheroid. Note that the frequencies plotted in this figure are squared and normalized by a factor shown in $y$-axis label of the plot. For any given $n$, we find that $\omega_r^2 \leq \Omega_0^2 \leq \omega_z^2$, and the equality holds only when $e=0$ or $e=1$ (for $r>a$). While the angular velocity and vertical epicyclic frequency decrease monotonically with radius for a given $e$ and $n$, the radial epicyclic frequency exhibits a local maximum for $n \leq 3$.

\begin{figure}
\begin{subfigure}[t]{0.5\textwidth}
         \centering
         \includegraphics[width=\textwidth]{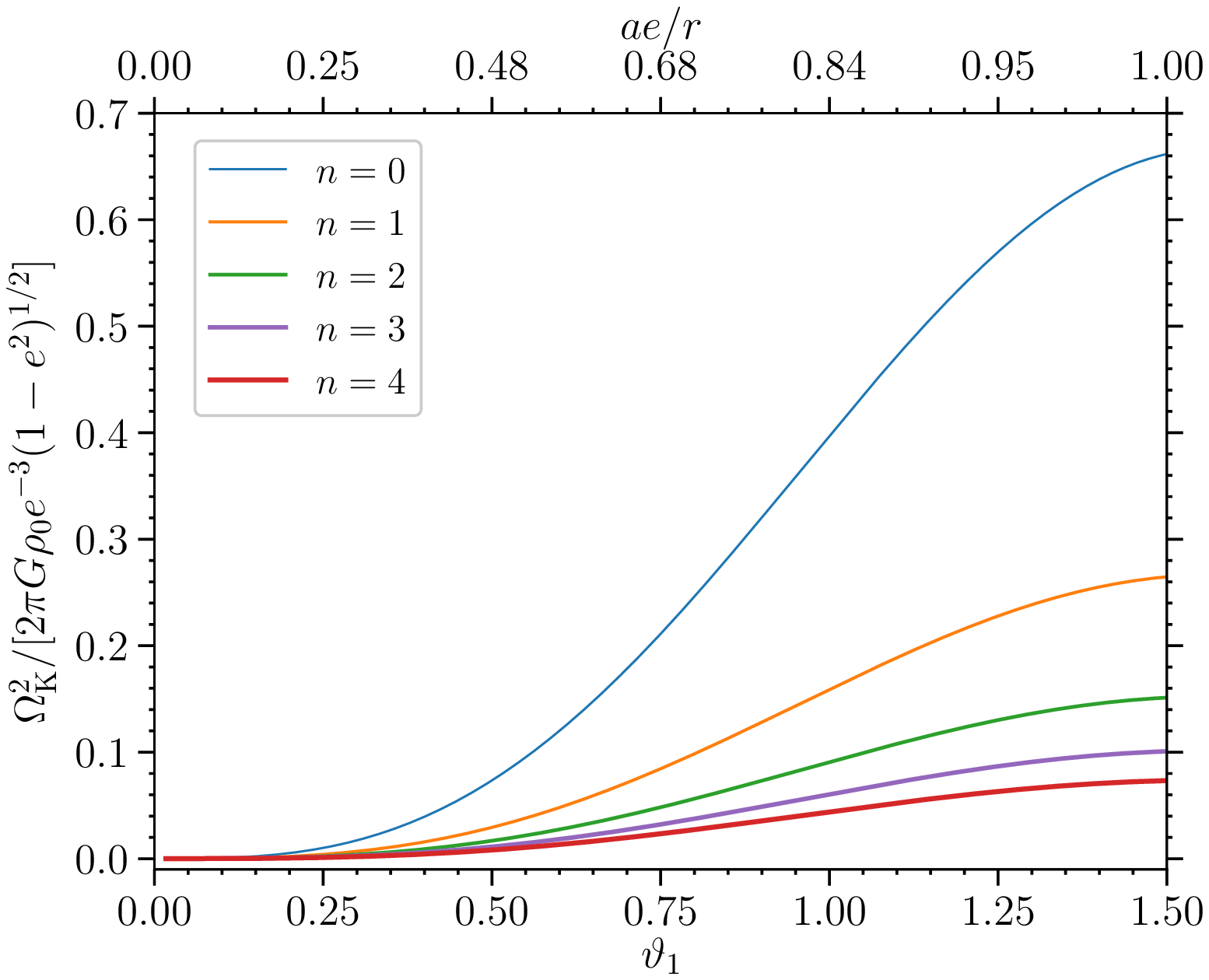}
\end{subfigure}
\begin{subfigure}[t]{0.5\textwidth}
         \centering
         \includegraphics[width=\textwidth]{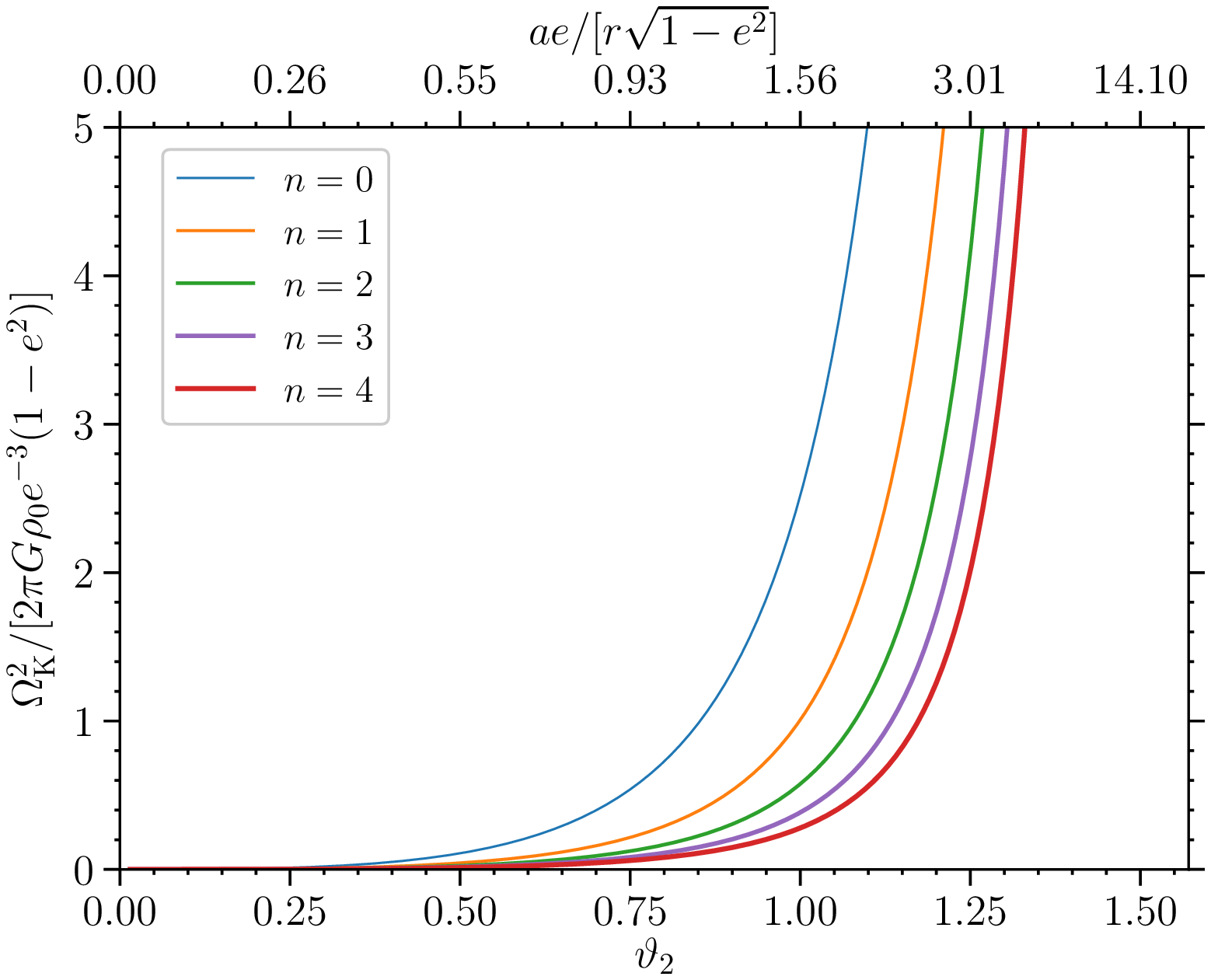}
\end{subfigure}
\caption{Square of Keplerian frequency is plotted as a function of $\vartheta_1$ ($\vartheta_2$) for oblate (prolate) spheroid in the top (bottom) panel for $n = 0,1,2,3,4$ represented by the curves with increasing line thickness, respectively.}
\label{fig:freq_Kep}
\end{figure}

In the case of a prolate spheroid, it is no longer possible to parameterize the frequencies in terms of $ae/r$ due to an additional term, $\sqrt{1-e^2}$, present in the definition of $\vartheta_2$. So instead, we plot the frequencies as a function of the compactified radius scaled with $\sqrt{1-e^2}$ as shown in the bottom panel of Fig.~\ref{fig:freq}. Note that the normalization factor for the frequencies is different in the prolate case, due to the difference in the value of $\zeta$ in Eq.~\ref{eq.omega_bin} and also the $(1-e^2)^{3/2}$ factor present in the $\alpha$-coefficients contribute. In contrary to the oblate case, $\omega_z^2 \leq \Omega_0^2 \leq \omega_r^2$ holds true for any given $n$, and once again the equality holds true when $e=0$ (for $r>a$)\footnote{For the prolate configuration, the radial epicyclic and orbital frequencies are discontinuous at $e=1$ for any $r>a$. It is an unphysical limit since it corresponds to no star at all, i.e., $M=0$. For the oblate configuration $M\rightarrow 0$ as $e\rightarrow 1$, so formally everything is continuous. In the prolate case, $M\rightarrow \infty$ as $e \rightarrow 1$, hence the ``discontinuity". The vertical epicyclic frequency goes to zero in the prolate case because the star resembles a cylinder (or wire) $e \rightarrow 1$.}.

Fig.~\ref{fig:freq} gives the impression that the frequencies for a prolate spheroid tend to infinity near the stellar surface because mathematically $\tan \vartheta_2$ can take values up to infinity. In reality, the frequencies approach a finite value as shown in the bottom panel of Fig.~\ref{fig:freq_e0p5_e0p9} due to the constraint $a/r<1$. Fig.~\ref{fig:freq_e0p5_e0p9} shows the radial profiles of frequencies around an oblate (top panel) and prolate spheroid (bottom panel) for a fixed $e$. For $e=0.5$ (shown in the left panel plots), the radial profiles of the frequencies for oblate and prolate spheroids are qualitatively similar. Quantitatively, the frequencies at a given radius around a prolate spheroid are higher than the oblate spheroid by a few to 10 per cent. At higher eccentricities, the radial profiles of frequencies look qualitatively different for an oblate and prolate spheroid, particularly the presence of a maximum in the radial epicyclic frequency in the former. It is very clear from this figure that the larger the eccentricity the larger is the deviation of the epicyclic frequencies from the orbital frequency.

It is easy to verify that when $e\rightarrow 0$ or $r/a \rightarrow \infty$, the frequencies approach the Keplerian value, i.e., $\Omega_0^2 = \omega_r^2=\omega_z^2 = GM/r^3$. This is valid for any value of n, since the mass of the ellipsoid for $n>0$ is given by 
\begin{equation}
    M =  4 \pi a^3 (1-e^2)^{\zeta} \rho_0 \int_0^1 (1-m^2)^n m^2 {\rm d}m
    \label{eq:mass}
\end{equation}

\begin{figure}
\begin{subfigure}[t]{0.5\textwidth}
         \centering
         \includegraphics[width=\textwidth]{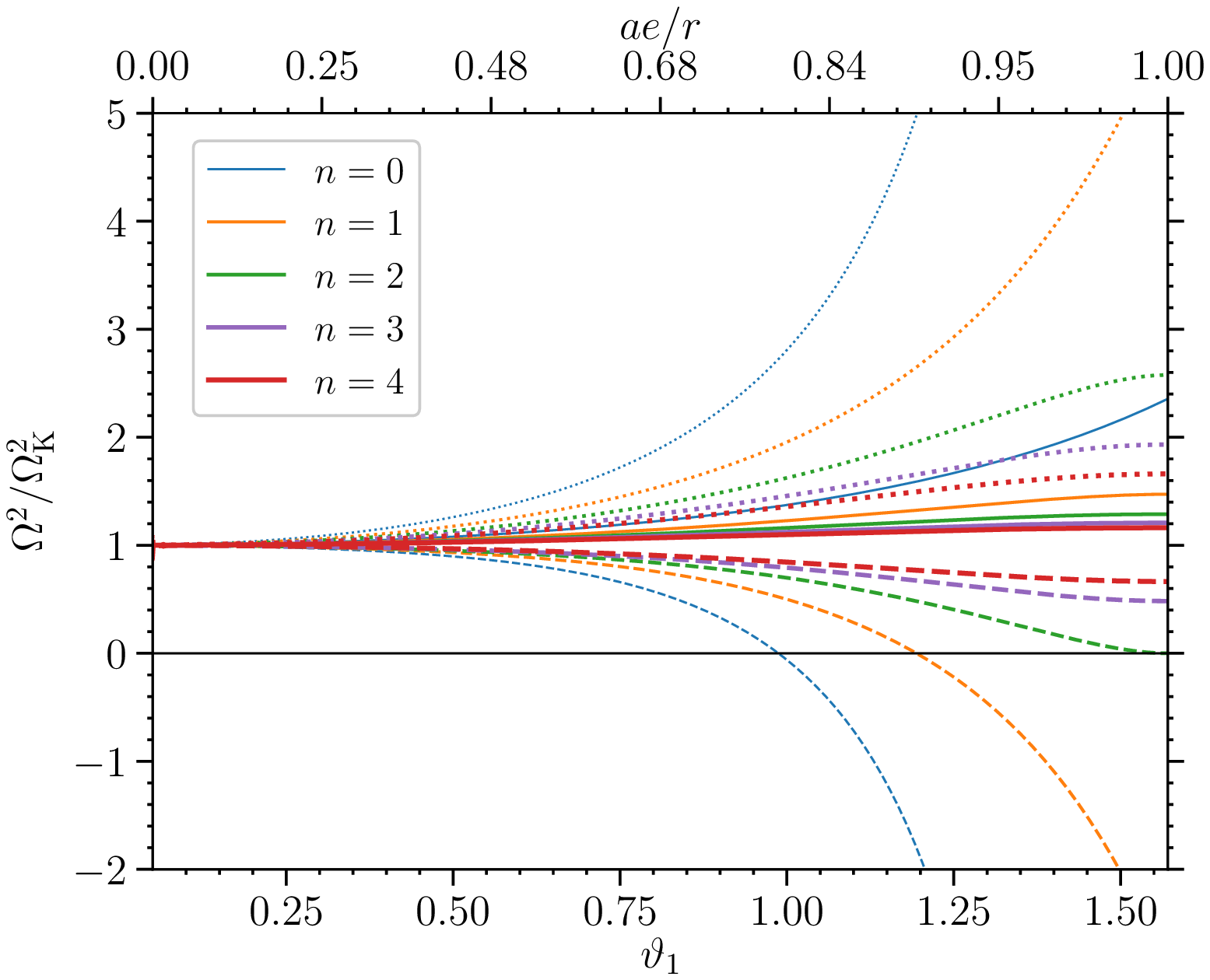}
\end{subfigure}
\begin{subfigure}[t]{0.5\textwidth}
         \centering
         \includegraphics[width=\textwidth]{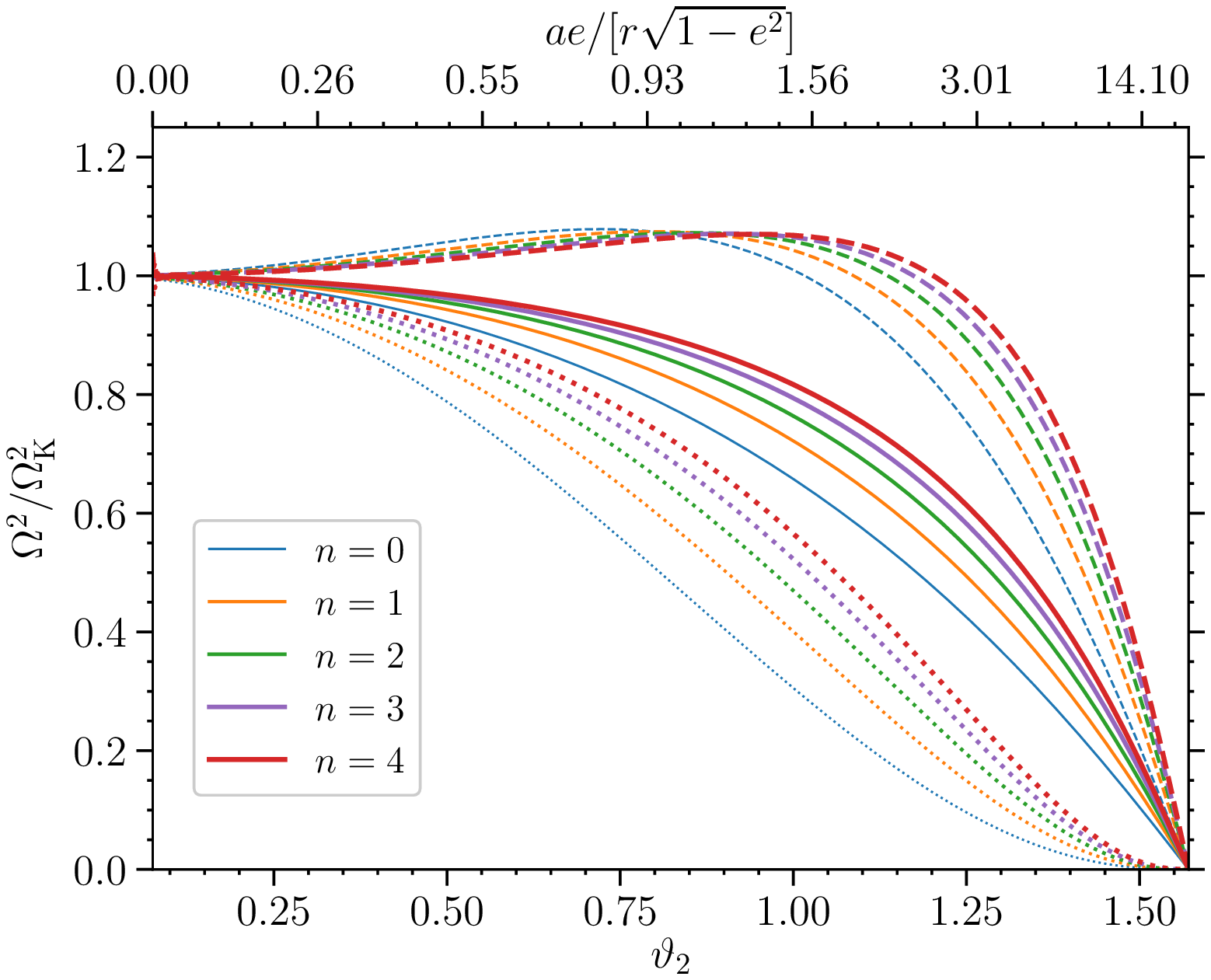}
\end{subfigure}
\caption{Squares of ratio of angular velocity (solid curves), radial epicyclic frequency (dashed curves) and vertical epicyclic frequency (dotted curves) to the Keplerian frequency is plotted as a function of $\vartheta_1$ ($\vartheta_2$) for oblate (prolate) spheroid in the top (bottom) panel for $n = 0,1,2,3,4$ represented by the curves with increasing line thickness, respectively.}
\label{fig:freq_Kep_ratio}
\end{figure}
With the help of Eq.~\ref{eq:mass}, we can summarize the results of the previous section as 
\begin{equation}
   \Omega^2/\Omega_{\rm K}^2 =  \frac{\sum_{j=0}^n (-1)^j \binom nj {\cal X}_j \left(\dfrac{r}{ae}\right)^{2j+3}}{\sum_{j=0}^n (-1)^j \binom nj \frac{2}{2j+3}}, 
\end{equation}
where
\begin{equation}
 \Omega_{\rm K}^2 = \frac{GM}{r^3} =  4\pi G\rho_0 (1-e^2)^{\zeta}\frac{a^3}{r^3} \int_0^1 (1-m^2)^n m^2 {\rm d}m,
 \label{eq:Kep}
\end{equation}
and
\begin{eqnarray}
 {\cal X}_j =\begin{cases}
   \alpha_j, & \text{for $\Omega^2 = \Omega_0^2$},\\
   (2j+4)\alpha_j+r\frac{{\rm d}\alpha_j}{{\rm d} r}, & \text{for $\Omega^2 = \omega_{r}^2$},\\
   \beta_j, & \text{for $\Omega^2 = \omega_{z}^2$}.\\
  \end{cases}
\end{eqnarray}
Note from Eq.~\ref{eq:Kep} that the Keplerian frequency is a now a function of $e$ and $n$, in addition to the radius. For a given $n$, the integral over $m$ gives a constant coefficient, and $\Omega_K^2$ varies as a third power of $\sin\vartheta_1$  and $\tan \vartheta_2$ for an oblate and prolate spheroid, respectively. Fig.~\ref{fig:freq_Kep} shows this dependence of Keplerian frequency on $\vartheta_1$ and $\vartheta_2$ for different values of $n$.

The top panel of Fig.~\ref{fig:freq_Kep_ratio} shows the ratio of angular and epicyclic frequencies to the Keplerian frequency for an oblate spheroid for the first few values of $n$. One can immediately see from this figure that for all $n$, as $e\rightarrow 0$ or $r\rightarrow \infty$, $ae/r \rightarrow 0$, at which the ratio goes to one. At a given radius, while the ratio of the radial epicyclic frequency to the Keplerian frequency decreases with $e$, the ratio of the orbital and vertical epicyclic frequencies to the Keplerian frequency increases with $e$. Note that all three frequencies decrease with the distance from the stellar surface. A similar plot for prolate spheroid is shown in the bottom panel of the same figure where the frequencies converge to the Keplerian value when $r\rightarrow \infty$ or $e\rightarrow 0$, as $ae/(r\sqrt{1-e^2}) \rightarrow 0$. Contrary to the oblate case, the ratio of the orbital and vertical epicyclic frequencies to the Keplerian frequency decreases with $e$ for a given radius, while the ratio of the radial epicyclic frequency increases with $e$ until it reaches a maximum somewhere near $e>0.65$ for $n\ge0$. One interesting finding is that for both the oblate and prolate spheroids, as $n$ increases, the orbital frequency tends towards the Keplerian value. This is due to the increase in the core density relative to the surface, which makes the star’s gravitational potential mimic a point-like source with the mass given by the core mass. While the epicyclic frequencies maintain the same trend in the oblate spheroid case, the radial epicyclic frequencies for different $n$ normalized with the Keplerian value exhibit a maximum for higher eccentricities in the prolate case. 

\subsection*{Presence of marginally stable orbit}
The existence of a marginally stable orbit is a special feature of general relativity. At this orbit ($r= r_{\rm ms}$), the specific angular momentum defined as $l(r) = \Omega_0(r) r^2$ has a minimum, which means that its first derivative vanishes at this orbit, i.e., $({\rm d}l/{\rm d} r)|_{r = r_{\rm ms}} = 0$. The radial epicyclic frequency also goes to zero at the marginally stable orbit, since $\omega_r^2 = r^{-3} {\rm d} l^2/{\rm d} r$. We find that some cases of rapidly rotating stars also harbour a marginally stable orbit even in Newtonian gravity.

In case of an oblate spheroid, Table~\ref{tab1} shows the location of the marginally stable orbit ($r=r_{\rm ms}$) in terms of the (compactified) radius, $ae/r_{\rm ms}$, for $n\leq 2$. When $r_{\rm ms} = a$, we obtain the lower limit on the possible values of eccentricity, $e_0$, for which a marginally stable orbit exists. In other words, for any value of $e>e_0$, the marginally stable orbit will lie separated from the surface of the star. Also, another way to look at all this is that for $1\geq e\geq e_0$, the circular orbits within $a<r<r_{\rm ms}$ are Rayleigh unstable because $\omega_{\rm r}^2<0$ for this range of radii. For the simplest case, $n=0$, of an oblate spheroid, we recover the root $ae/r_{\rm ms} = 0.83458318$ for Maclaurin spheroid \citep{WR2013}. Note that for $n \geq 3$, the derivative is always positive and never approaches zero for any value of $ae/r<1$ (as can be seen in Fig.~\ref{fig:freq}). This implies that for $n\geq 3$, no matter how rapidly the star may rotate, all circular orbits are stable. 

In the case of a prolate spheroid, as one can see in Fig.~\ref{fig:freq}, $\omega_r^2$ is a monotonically decreasing function of radius, for any given $e$. Therefore, a marginally stable orbit does not exist around a prolate spheroid. It is not entirely surprising that we do not find a marginally stable orbit in the prolate case because the coefficient of the quadrupole term of the potential in the prolate case bears an opposite sign to that of the oblate configuration. The coefficient should be negative to make it possible for $\omega_r$ to vanish at a certain radius and while this condition is satisfied in the oblate case, we find that the coefficient is always positive in the prolate case. Another check would be to express $\omega_r^2 = (2 \Omega_0/r) {\rm d}l/{\rm d}r$ and since $\omega_r^2 > \Omega_0^2$ in the prolate case, this implies that $ {\rm d}l/{\rm d}r > r \Omega_0/2$. If $ {\rm d}l/{\rm d}r$ goes to zero (which happens at the marginally stable orbit), then $\Omega_0$ is required to be negative. In the oblate case, this is no problem because $\omega_r^2 <\Omega_0^2.$

\begin{table*}
\centering
\caption{Location of the marginally stable orbit (in terms of compactified radius, $ae/r_{\rm ms}$) for different values of $n$ in case of an oblate spheroid}
\begin{tabular*}{0.7 \textwidth}{cccc}
\hline
n  & 0 & 1 & 2                 \\
\hline
$ae/r_{\rm ms}$ or $e_0$ (when $r_{\rm ms}=a$) & 0.8345831778601294 & 0.9299556854562332 & 0.999999999999999\\
\hline
\end{tabular*}
\label{tab1}
\end{table*}

A general requirement for a stable, thin accretion disc to exist around an accreting object is that the angular momentum increases monotonically with radius, i.e., ${\rm d} l^2/{\rm d} r > 0$ -- well known \textit{Rayleigh instability} criterion, and for accretion to be driven by magnetorotational instability (MRI) in a magnetized, differentially-rotating medium, we need ${\rm d} \Omega_0^2/{\rm d} r < 0$\footnote{Typically, the boundary layer between the star and the accretion disc has ${\rm d} \Omega_0^2/{\rm d} r > 0$, and even so, acoustic waves driven by the shear instabilities can potentially transport the angular momentum, making it possible for the accretion to occur in the boundary layer \citep{BR2012, BRS2013, CRP2022}. }; for more details, see chapters 7 and 22 of \citet{shu1992}. We find that the accretion discs around an oblate spheroid are Rayleigh stable at all radii beyond the marginally stable orbit for all possibilities of $e$ and $n$, while the same holds true for discs around a prolate spheroid for all $r>a$. It is clear from Fig.~\ref{fig:freq} that the MRI criterion for accretion in a magnetized disc around an oblate/prolate spheroid is satisfied.

\subsection*{Similarities to Kerr metric}
For the Newtonian $1/r$ potential, irrespective of the motion of the object, both the radial and the vertical epicyclic frequencies of particles orbiting a spherically symmetric (rotating or non-rotating) object are equal to the angular frequency. This is no longer the case in general relativity, where the radial epicyclic frequency is smaller than the angular frequency ($\omega_r<\omega_z = \Omega_0$) for a non-rotating object that has a spherically symmetric metric. This degeneracy in the vertical epicyclic and angular frequency is broken for a rotating black hole (which has an axisymmetric metric) and depending on the direction of rotation, $\omega_z$ is either less than $\Omega_0 $ (like for a prograde Kerr orbit) or greater than $\Omega_0 $ (like for a retrograde Kerr orbit). Interestingly, axially symmetric objects in Newtonian gravity can also break the degeneracy among these frequencies~\citep{ABG02}. As shown in \citet{WR2013}, Maclaurin spheroids, which are oblate spheroids, have $\omega_r^2<\Omega_0^2<\omega_z^2$, similar to a retrograde Kerr orbit. In this paper, we generalized this relation for any $n$. Now one may anticipate that the prolate spheroid should be of the opposite nature, and indeed in this case we have $\omega_z^2 < \Omega_0^2$ similar to a prograde Kerr orbit. However, this implies that $\omega_r^2>\Omega_0^2$ according to Eq.~\ref{eq:relation}, which is not true for a prograde Kerr orbit.

\subsection*{Frequency ratios}
\begin{figure}
\begin{subfigure}[t]{0.5\textwidth}
         \centering
         \includegraphics[width=\textwidth]{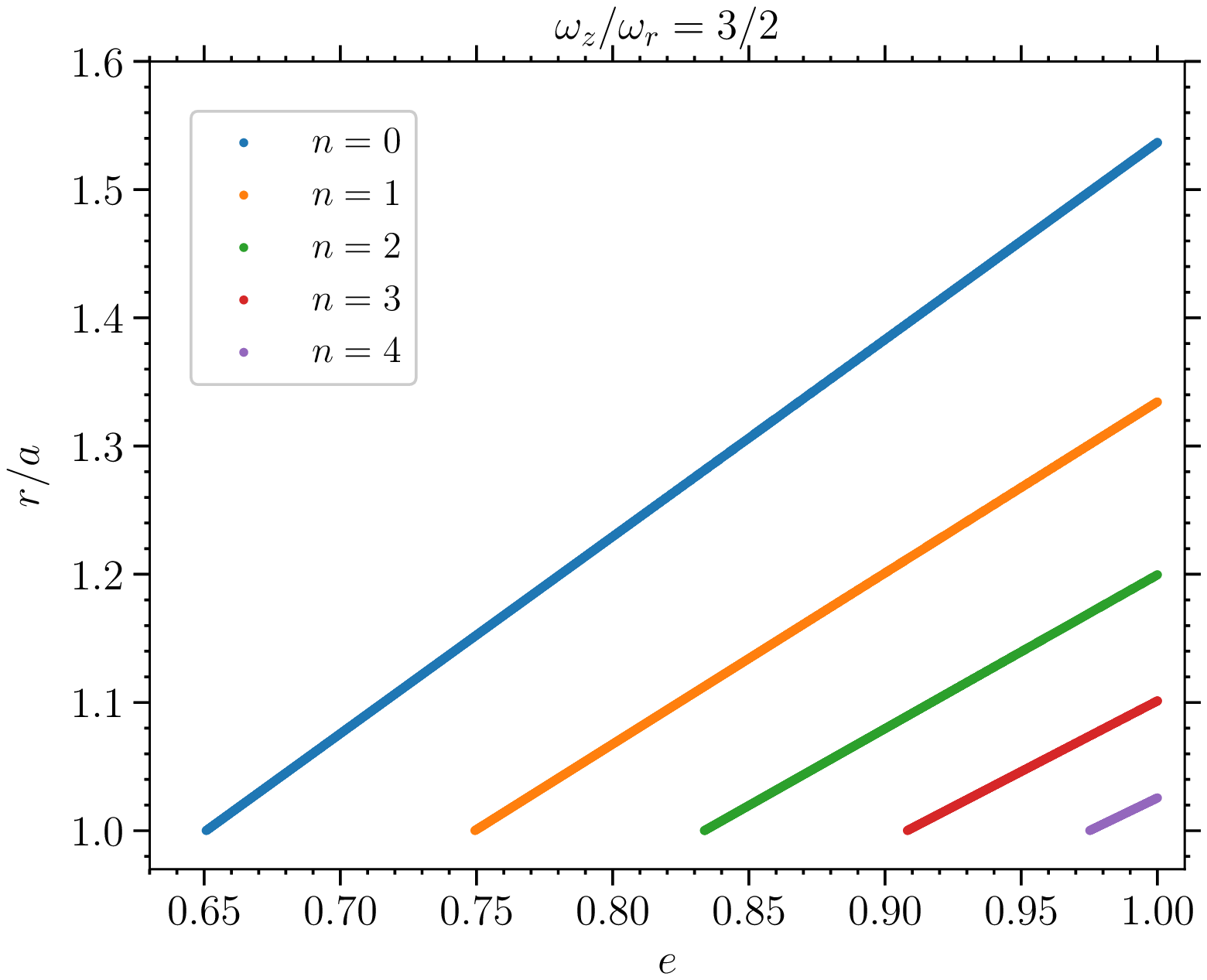}
\end{subfigure}
\begin{subfigure}[t]{0.5\textwidth}
         \centering
         \includegraphics[width=\textwidth]{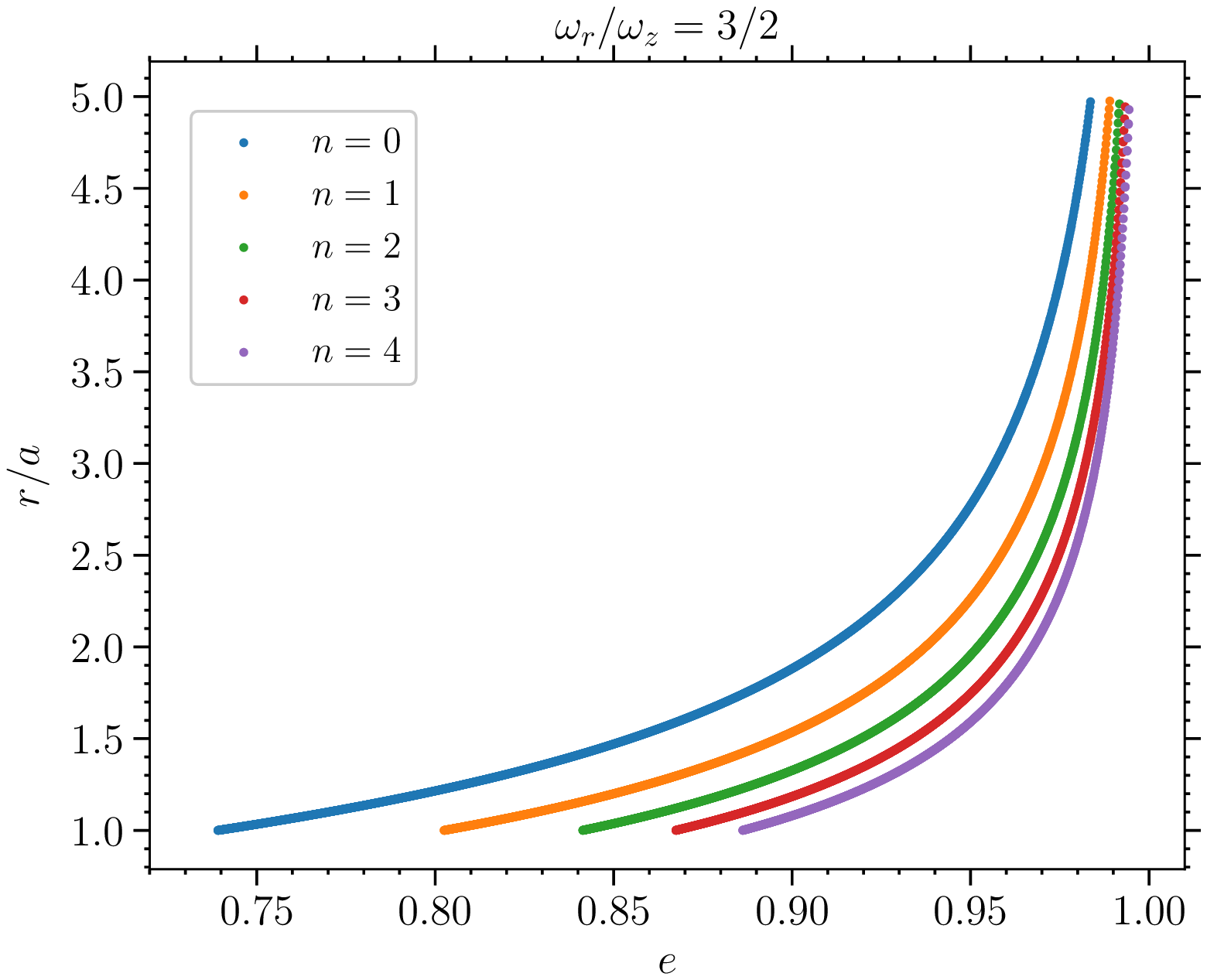}
\end{subfigure}
\caption{The locii of radius at which $\omega_z/\omega_r = 3/2$ for different values of $e$ and $n$ is shown in the top panel for an oblate spheroid. The same, but instead $\omega_r/\omega_z = 3/2$, is shown in the bottom panel for a prolate spheroid.}
\label{fig:ratio}
\end{figure}

The ratio of the epicyclic frequencies has great relevance to observations. In the case of the black hole and neutron star systems, the high-frequency QPOs are often found in pairs in the ratio of $3:2$ and various models have associated these frequencies with different combinations of angular/ epicyclic frequencies or some kind of resonance between them \citep{KA2005}. In the case of white dwarfs, the frequencies of the dwarf novae oscillations are reported in the ratio of $1:2:3$. However, due to the nature of the identity in Eq.~\ref{eq:relation}, it is not possible to have the ratio of the angular and epicyclic frequencies in the ratio of $1:2:3$. For example, in an oblate spheroid we have $\omega_r <\Omega_0<\omega_z$, so if $\Omega_0 = 2\omega_r$, then $\omega_z$ becomes $\sqrt{7}\omega_r$. So these frequencies cannot be associated with the dwarf novae oscillations. However, it is possible that the epicyclic frequencies could assume an integer ratio, like 3:2 at a particular radius for a given $e$. Fig.~\ref{fig:ratio} shows the possible locations from an oblate-spheroid (top-panel)/ prolate-spheroid (bottom panel) shaped star at which the vertical and radial epicyclic frequency are in $3:2$ ratio. With increasing $n$, only high eccentricities can produce the radial and vertical oscillations of orbit in a $3:2$ ratio for a given $r>a$. Also, note that at the marginally stable orbit where $\omega_r^2 = 0$, Eq.~\ref{eq:relation} implies that $\omega_z(r_{\rm ms}):\Omega_0 (r_{\rm ms}) = 1:\sqrt{2}$. 

\subsection*{Relevance to astrophysical sources}
Fig.~\ref{fig:freq_Kep} gives us an idea of how the frequencies scale linearly with the mass of the star and are inversely proportional to the radius of the star. For a system like \textit{SS Cyg} which has $M \sim 0.95~M_{\odot}$ and $a \sim 5.5\times 10^8\,{\rm cm}$ \citep{GGRP83, HSHS17}, assuming it to be oblate-shaped, the typical range of the orbital frequencies inferred at the surface ranges between $0.14$~Hz (Keplerian value) to $0.21$~Hz. As $e\rightarrow 1$, the vertical epicyclic frequencies can go higher than $5$~Hz and the radial epicyclic frequency takes values between $0.14$~Hz to zero for $n\leq2$. These values are comparable to the observed QPO frequencies of around 0.1~Hz observed in the EUV and soft X-ray flux of \textit{SS Cyg}~\citep{M1996, M2002} for lower values of $e$.

The discussed epicyclic frequencies in this paper could be relevant to real astrophysical objects like protoplanets or rapidly rotating stars/white dwarfs. Main sequence stars, particularly of higher mass tend to become rapidly rotating, oblate spheroids after undergoing a partial tidal disruption event (TDE) and the density profile of the remnant star is comparable to Eq.~\ref{eq.rho} for $n \sim 13$ \citep{RKP20}. The eccentric nature of the partially disrupted star in TDEs is a transitional phenomenon, but if rapid variability could be observed from such stars, it could reveal potential information about the interior of the star. Although at this point, we cannot make any direct comparison to the observations of the aforementioned astrophysical objects, our analytical expressions can only stand as predictions for possible future observations. Lastly, with the knowledge of the radial profile of the angular velocity, one can study the dynamical evolution of a geometrically thin disc around a prolate/oblate spheroid similar to the work by \citet{MV2015} done for a Maclaurin spheroid. Further, one can also compute the energy spectra from the disc, and we plan to carry out this work in the future. 

Theoretical studies suggest that the global toroidal magnetic fields \citep{CF1953, W61} or meridional flows within the star~\citep{FE2014} can potentially deform the star into prolate shape, but there exists no observational evidence yet for a prolate-shaped star. Detection of gravitational waves opened a new window for multi-messenger astronomy, inspiring several works in the last decade to study the deformation of neutron stars due to the magnetic fields~\citep{CR2013} or internal flows. Rough estimates of the eccentricities caused by the deformation due to the magnetic fields are around $0.2$ ~\citep{LJ2009}, while the maximal meridional circulation velocities can cause deformation at the surface to about one percent~\citep{BSM2011}.
\section{Summary}
\label{section5}
Previous studies have shown that the degeneracy in the epicyclic frequencies and orbital frequency of orbits in $1/r$ Newtonian potential can be broken by introducing the rotation of the star. However, these studies assume a Maclaurin spheroid; an oblate-shaped spheroid with uniform density. In this paper, we extend this study to both the oblate and prolate configurations of a rotating star, allowing for the star to have a non-uniform density. We assume a power-law profile for the density distribution inside the star, characterized by the power-law exponent, $n$. The explicit analytical expressions for the three important characteristic frequencies -- angular, radial epicyclic and vertical epicyclic frequencies -- around a rotating ellipsoid that has non-uniform density are given by Eqns.~\ref{eq.omega_bin},~\ref{eq:omegar_exp}, and~\ref{eq.omegaz}, respectively, with the parameter $\zeta = 1/2 (-1/2)$ for an oblate (prolate) spheroid. We find that the nature of the discussed characteristic frequencies differs for an oblate and prolate spheroid.

For any $n$, the orbits around an oblate spheroid are similar to the retrograde Kerr-like orbits in the respect that $\omega_r<\Omega_0<\omega_z$. On the contrary, the orbits around a prolate spheroid share similarity with prograde Kerr orbits in having $\omega_z<\Omega_0$. For an oblate spheroid, a marginally stable orbit can exist outside the star (just like for relativistic orbits) for all values of $e>e_0$ valid only for $n=0,1$~and~$2$ (see Table~\ref{tab1}). The marginally stable orbit does not exist for a prolate spheroid. This comes from the fact that the coefficient of the quadrupole term, which has to be negative for $\omega_r$ to vanish, is positive in the prolate case. We find that with increasing $n$, the three characteristic frequencies, both in the prolate and oblate case, tend towards the Keplerian value. This is due to the increase in the core density relative to the surface. 

The dwarf novae oscillations in the cataclysmic variables are often observed in $1:2:3$ ratio. Eq.~\ref{eq:relation} rules out the possibility that the ratio of $\omega_r:\Omega_0:\omega_z$ can be $1:2:3$. However, as we briefly discussed in the paper, these frequencies fall in the observed range of QPO frequencies in Cataclysmic variables. Further, we showed that the ratio of the epicyclic frequencies can occur in the integer ratio of 3:2 for certain combinations of $e$ and $r/a$. (see Fig~\ref{fig:ratio}).

The frequencies discussed in this paper can be a viable model for the QPOs observed in the cataclysmic variables. They are also relevant for accretion discs around any rotating, compact stars. At a given radius, and for a given ellipticity, we find that the frequencies of a prolate star are higher than the oblate star by a few per cent. Also, at a given radius, the frequencies decrease with $n$. Such properties of these frequencies can be used as a probe to study the properties of the compact star. \citet{GKSW2014} showed that the properties of orbital and epicyclic frequencies in the rotating strange quark stars are a result of the interplay of competing general relativity and Newtonian effects. It would be interesting to study how the orbital and epicyclic frequencies computed here compare with the frequencies of the rotating strange quark stars for different rotation rates. We plan to pursue this in the future.

\section*{Acknowledgements}
The author is grateful to the referee for constructive comments that improved the quality of the paper. The author is greatly indebted to W{\l}odek Klu{\'z}niak for all the discussions and encouragement to publish this study. The author also thanks Eiichiro Komatsu, Bhupendra Mishra, and Chandrasekhar Saraf for all the useful discussions and comments. This research is partly supported by the Polish National Science Center (NCN) grant 2018/29/N/ST9/02139.

\section*{Data Availability}
No new data were generated or analysed in support of this research.


\bibliographystyle{mnras}
\bibliography{hetero_spheroid} 


\bsp	
\label{lastpage}
\end{document}